\documentclass[conference]{IEEEtran}
\usepackage{cite, epsfig, mathptmx, times, amssymb, amsmath, color}
\usepackage{subcaption}

\usepackage{graphicx,booktabs,soul,color,graphics,epsfig,mathptmx,moreverb,ifpdf,cancel,amsthm,framed,dblfloatfix}
\usepackage{gensymb,graphicx,url, multirow,kantlipsum}
\usepackage{fancyhdr}
\usepackage{tabularx,caption}
\usepackage{mhchem}
\usepackage{siunitx}
\usepackage{multirow}
\usepackage{tabularx}
\newcolumntype{Z}{>{\raggedright}X}

\usepackage{lipsum}

\makeatletter
\def\ps@IEEEtitlepagestyle{%
	\def\@oddfoot{\mycopyrightnotice}%
	\def\@evenfoot{}%
}

\def\ps@headings{%
	\def\@oddhead{\mbox{}\rightmark \textbf{\emph{2023 IEEE PES/IAS PowerAfrica}} \hfil}%
	\def\@evenhead{}%
}
\def\mycopyrightnotice{%
	{ \footnotesize 979-8-3503-3755-6/23/\$31.00 ~ \copyright 2023 IEEE\hfill}
}

\makeatother

\DeclareRobustCommand*{\IEEEauthorrefmark}[1]{%
	\raisebox{0pt}[0pt][0pt]{\textsuperscript{\footnotesize #1}}%
}

\begin{document}
%
\title{Lead-acid battery lifetime extension in solar home systems under different operating conditions}

\author{\IEEEauthorblockN{\IEEEauthorblockN{
			R. Perriment\IEEEauthorrefmark{1}, V. Kumtepeli\IEEEauthorrefmark{1}, M.D. McCulloch\IEEEauthorrefmark{1},
			D.A. Howey\IEEEauthorrefmark{1}
		}
		\IEEEauthorblockA{
			\IEEEauthorrefmark{1}Department of Engineering Science, University of Oxford, OX1 3PJ, Oxford, United Kingdom \\
			}}}

\maketitle

\begin{abstract}
Solar home systems (SHS) provide low-cost electricity access for rural off-grid communities. Batteries are a crucial part of the system, however they are often the first point of failure due to shorter lifetimes. Using field data, this work models the degradation of lead-acid batteries for different SHS use-cases, finding the dominant ageing mechanisms in each case. Corrosion is the dominant ageing mechanisms in all cases apart from the highest use case. This is caused by extended time at high state of charge (SOC) and hence high voltage. A new voltage control scheme is proposed for one of the use cases dominated by corrosion, whereby the number of days between full recharges varies depending on the degradation mechanisms the battery experiences. Simulating the new voltage control scheme yields a 25\% increase in battery lifetime whilst ensuring no loss of load to the user.
\end{abstract}


\section{Introduction}
Sustainable Development Goal 7 highlights the importance of sustainable, reliable and affordable energy access for all \cite{GoalAffairs}, yet there are still more than 750 million people lacking access to electricity globally, of which 75\% live in Sub-Saharan Africa. SHS are an important primary step for enabling energy access, especially in rural locations with no grid-connection. However, batteries are a disproportionately expensive part of these systems, at over 80\% of total cost \cite{Charles2019SustainableContext} and are often the first point of failure. Therefore, increasing battery lifetime is crucial for reducing costs.

Lead-acid batteries are a mature technology with a lower upfront cost than lithium-ion, and they therefore remain common in SHS in rural low- and lower-middle- income countries (LMIC). Lead-acid batteries have the additional benefit of having closed-loop recycling potential with infrastructure for this already established in many LMICs, furthering the local economy and reducing environmental impact \cite{Charles2019SustainableContext}. However, lead-acid batteries have a generally shorter lifetime \cite{PaulAyengo2018ComparisonBatteries} and finding ways to improve this, including for systems already in the field, is crucial in order to maintain and expand energy access. 

Different charging strategies can significantly impact battery lifetime \cite{Lavety2020EvaluationBattery}. The typical charging strategy for lead-acid batteries in solar applications is the three-stage-charging-cycle (TSCC)\cite{Bogno2017ImprovementSystems}. TSCC begins with a `bulk' charging phase at constant current until an upper voltage limit is reached. Then the `absorption' phase follows, where the current is reduced to maintain a constant voltage. Once the battery is fully charged, the voltage is reduced to a float voltage, which is maintained until the battery is discharged, and the cycle starts again \cite{Bogno2017ImprovementSystems}. This charging strategy is common in off-grid lead-acid systems -- however, the voltage limits used are non-trivial and can largely impact the battery state of health (SOH) \cite{J.Badeda2017AdaptiveApplications}. 

Svoboda \emph{et al.}  \cite{Svoboda2007OperatingSystems} define a set of stress factors for lead-acid battery degradation: charge factor, Ah throughput, highest discharge rate, time between full charge, time at low SOC and partial cycling. These stress factors can be calculated from battery time series data, namely the current, voltage, and temperature \cite{Lujano-Rojas2016OperatingMicrogrids}. Additionally, the time series data can be an input to battery degradation models to simulate the degradation over time. There are many battery models in the literature, from simple linear cycle models \cite{Dufo-Lopez2014ComparisonSystems} to equivalent circuit models \cite{Schiffer2007ModelSystems} to electrochemical models derived from physical first principles \cite{Sulzer2019FasterModel}.

A common model in literature is the Schiffer model \cite{Schiffer2007ModelSystems}. This considers the chemical processes involved in lead-acid battery ageing; corrosion, sulphation, degradation of the active material and gassing. It is generally accepted to be the most accurate systems-level model \cite{Dufo-Lopez2014ComparisonSystems, Lujano-Rojas2016OperatingMicrogrids, Bashir2017LifetimeSystems} while having the additional benefit of categorising the degradation as either corrosion or active mass degradation, which can be useful to identify causes of failure. While the Schiffer model does not directly model the electrochemical mechanisms within the battery, the more qualitative approach is based on understanding of the mechanisms and allows a simpler implementation over long time series.

This work aims to combine knowledge from literature on charging strategies and operating conditions of lead-acid batteries with field data on how the batteries in the SHS are used. The field data is from an African based SHS provider BBOXX. Initially, different degradation mechanisms are modelled for different battery use cases. Then a new voltage control method is proposed for a specific use case, which considers how the lifetime can be increased without impacting the SHS user experience.

\section{Solar Home System Design}
\begin{figure}
    \centering
    \includegraphics[width=0.9\linewidth]{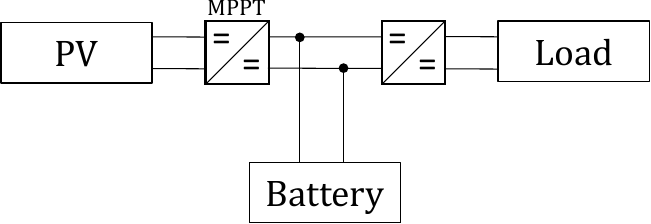}
    \caption{Power flow schematic for BBOXX SHS}
    \label{fig:system}
\end{figure}

The BBOXX SHS consist of a 50 Wp solar panel and a 12 V, 20 Ah valve-regulated lead-acid (VRLA) battery. The SHS are used to meet the load demand, made up of low power ($<$8 W) appliances including lighting, phone charging, fans and entertainment devices such as TVs and radios. Fig. \ref{fig:system} demonstrates the power flows in the system.

BBOXX use internet of things technology to collect time series data from the SHS (including battery current, voltage, and temperature), while allowing remote control of the SHS, such as changing voltage limits or turning the SHS on or off \cite{Aitio2021PredictingLearning}.

\subsection{SHS use cases}
There is significant variation in the load demand of different households, and therefore it is essential to consider multiple different load demands when trying to improve the battery performance -- a one size fits all approach does not suffice. Previous work was done on clustering the load demand of 1,077 BBOXX's SHS \cite{Kumtepeli2023FastC++}. The results were 4 different clusters, each with representative centroids. These centroids are used as simulation inputs in this work. The clusters can be archetypally defined as:

\begin{enumerate}
    \item Regular high use
    \item Regular moderate use
    \item Regular low use
    \item Infrequent use, with long periods of non-use
\end{enumerate}

\section{Battery Model}

The battery model used is based on the Schiffer weighted Ah-throughput model \cite{Schiffer2007ModelSystems}, where the full model is detailed. Some modifications have been made to improve the model. The open circuit voltage (OCV) and positive terminal OCV are calculated using the method proposed in \cite{Sulzer2019FasterModel}:
\begin{align}
\label{batb}
    U(t) &= 1.92 + 0.15y + 0.06y^{2} + 
    0.07y^{3} + 0.03y^{4} \\
      \label{pos_v}
    U_{\text{P}}(t) &= 1.628 + 0.074y + 0.033y^{2} + 0.043y^{3} + 0.022y^{4}
\end{align}

Where $y$ is the 10-base logarithm of acid molality and can be calculated from the acid molarity ($c$):
\begin{align}
    y(t) &= \log_{10} \left(\frac{c(t)V_{\text{w}}}{(1-c(t)V_{\text{e}})M_{\text{w}}} \right)\\
        c(t) &= c_{\text{max}}+\frac{C_{\text{N}}}{{\text{F} A_{\text{elec}}}\left(\text{SOC(t)}-1\right)}
\end{align}
where $C_{\text{N}}$ is the nominal battery capacity, $\text{F}$ is the Faraday constant, $A_{\text{elec}}$ is the electrolyte volume (=$1.43 \times 10^{-4}$ m$^{3}$), $V_{\text{w}}$ is the molar volume of \ce{H2O} (=17.5 cm$^{3}$ mol$^{-1}$), $V_{\text{e}}$ is the molar volume of \ce{H2SO4} (=45 cm$^{3}$ mol$^{-1}$) and $M_{\text{w}}$ is the molar mass of \ce{H2O} (=18 g mol$^{-1}$).

The voltage is calculated based on the modified Shepard equations as in \cite{Schiffer2007ModelSystems}:
\begin{equation}
    V(t) = U(t) + b_{0}(t)\frac{I(t)}{C_{\text{N}}} + b_{0}(t)b_{1}\frac{I(t)}{C_{\text{N}}}\frac{\text{SOC(t)}}{1-\text{SOC(t)}}
\end{equation}
where $I$ is the battery current ($I(t) > 0$ is charging), $b_{0}$ is the aggregated internal resistance, $b_{1}$ is the charge-transfer overvoltage coefficient. 

The SOC is calculated using a coulomb counting approach \cite{Schiffer2007ModelSystems}
\begin{equation}
    \text{SOC(t)} = \int \frac{I(t) - I_{\text{gas}}(t)}{C_{\text{N}}} \,dt
\end{equation}
Using time series data causes SOC drift over time, due to accumulation of small reading inaccuracies or missing data points. In addition, coulomb counting causes SOC drift over a battery's lifetime regardless of data inaccuracies due to changes in battery capacity -- a 2 Ah discharge will lead to a larger SOC decrease later in the battery's life when its capacity is reduced. Therefore, the SOC is fixed at any point with a current less than 0.01 A by inverting the OCV. Additionally, whenever float charge is reached, SOC is assumed to be 100\%.

The gassing current is dependent on voltage and temperature, and can be modelled using a Tafel approximation \cite{Schiffer2007ModelSystems}:
\begin{equation}
\label{bate}
    I_{\text{gas}}(t) = I_{\text{gas,0}} \exp{(c_{\text{V}}(V(t)-V_{\text{gas,0}})+c_{\text{T}}(T(t)-T_{\text{gas,0}}))}
\end{equation}
where $I_{\text{gas,0}}$ is a normalised gassing current (=0.017 A), $c_{\text{V}}$ is the voltage coefficient (=0.183 V$^{-1}$), $c_{\text{T}}$ is the temperature coefficient (=0.06 K$^{-1}$), $V_{\text{gas,0}}$ is the nominal voltage (=13.38 V) and $T_{\text{gas,0}}$ is the nominal temperature (=298 K).

\subsection{Corrosion}

Corrosion is assumed to take place on the positive electrode, hence the positive terminal voltage is used, which assumes half of the overpotentials are associated with the positive terminal \cite{Schiffer2007ModelSystems}. 
\begin{equation}
\label{cb}
    V_{\text{P}} = U_{\text{P}} + \frac{1}{2}\left(b_{0}(t)\frac{I(t)}{C_{\text{N}}} + b_{0}(t)b_{1}\frac{I(t)}{C_{\text{N}}}\frac{\text{SOC(t)}}{1-\text{SOC(t)}}\right)
\end{equation}

\begin{figure}
    \centering
    \includegraphics[width=0.95\linewidth]
    {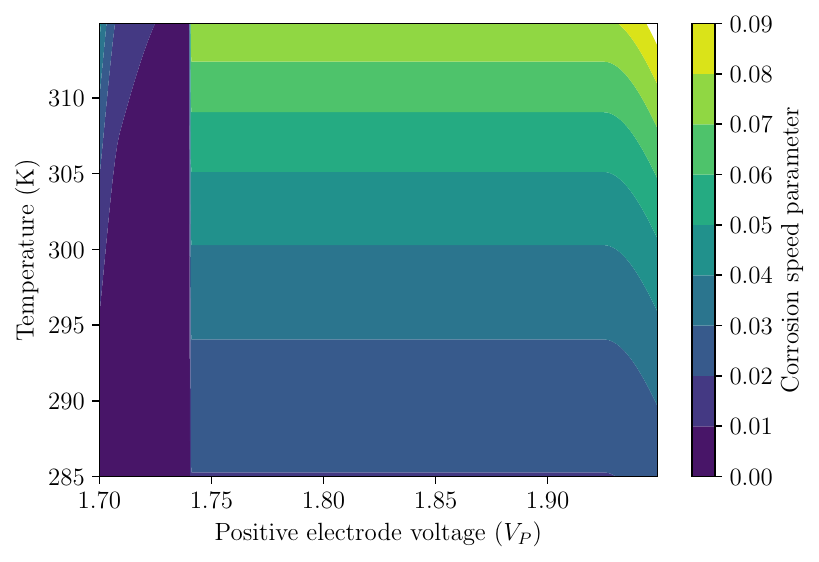}
    \caption{Variation in the corrosion speed parameter with battery temperature and positive electrode voltage.}
    \label{fig:ks}
\end{figure}

When at high SOC, the anodic potential is very high which accelerates corrosion of the lead grid in the positive electrode into lead oxides \cite{Ruetschi2004AgingBatteries}, reducing the conductivity of the grid and causing a loss of active material, thus decreasing the battery capacity. The corrosion layer thickness, $W$, is calculated as in \cite{Schiffer2007ModelSystems} by first calculating the corrosion speed parameter. The corrosion speed parameter varies with $V_{\text{P}}$ and temperature as depicted in Fig. \ref{fig:ks}.
\begin{equation}
\label{ce}
    C_{\text{corr}}(t) = C_{\text{corr,limit}} \frac{W(t)}{W_{\text{limit}}}
\end{equation}
where $C_{\text{corr,limit}}$ is the battery capacity reduction at end of life (EOL) due to corrosion and $W_{\text{limit}}$ is the corrosion layer thickness at EOL based on the datasheet float lifetime.

\subsection{Active mass degradation}

Active mass degradation occurs as a result of the battery cycling process. Active mass degradation is modelled by calculating a weighted number of cycles ($Z_{\text{w}}$) depending on SOC and discharge current ($I_{\text{d}}$) \cite{Schiffer2007ModelSystems}. As the batteries in this work are gel VRLA, acid stratification is not considered \cite{Wagner2001ChargeApplications}.
\begin{equation}
\label{amb}
    Z_{\text{w}}(t) = \frac{1}{C_{\text{N}}} \int_{0}^{t} I_{\text{d}}(\tau) f_{\text{SOC}}(\tau) d\tau
\end{equation}
where $f_{\text{SOC}}$ is the SOC factor, which depends upon the time  since the last full charge and the minimum SOC reached since the last full charge.
The capacity loss due to active mass degradation is calculated by
\begin{equation}
\label{ame}
    C_{\text{deg}}(t) = C_{\text{deg,limit}} \exp \left[-5 \left(1- \frac{Z_{\text{w}}(t)}{Z_{\text{N}}}\right)\right]
\end{equation}
where $C_{\text{deg,limit}}$ is the battery capacity reduction at EOL due to active mass degradation and $Z_{\text{N}}$ is the datasheet nominal number of cycles util EOL. The total battery degradation is
\begin{equation}
\label{tot}
    C(t) = C_{\text{corr}}(t)  + C_{\text{deg}}(t)
\end{equation}

In VRLA batteries, failure is actually a result of conductance decrease (i.e., resistance increase) rather than decreased thermodynamic capacity due to reduced lead/lead dioxide activity \cite{Bouet1981AnalysePlomb}. Since capacity and conductance are strongly correlated, it is equivalent to consider either of them, so this paper focuses on capacity fade.

\subsection{Voltage control}
\label{vc}
The voltage control system used by BBOXX SHS is TSCC. This work is focused on prolonging the life of systems already in the field, and therefore hardware updates are not considered. However, over the air updates are possible for systems in the field, so the voltage limits can be varied.

Previous work on lead-acid battery operation finds that a full recharge is only necessary once every 8 days to prevent sulphation \cite{Svoboda2007OperatingSystems}. The BBOXX systems get a full recharge on 95\% of days (calculated from the sample of 1,077 SHS). This means they are at greater risk of corrosion due to more time at high voltages. This work proposes using variable voltage limits depending on the degradation mechanism impacting the battery the most. In the proposed voltage control strategy, a full recharge is not enforced every day, and the more degradation caused by corrosion, the less frequently the battery fully recharges. Fig. \ref{fig:ks} shows that in the region where $V_{\text{P}} < 1.74~\text{V}$ the corrosion speed parameter is smaller. This corresponds to a battery voltage of 12.8V and SOC of 89\% when the OCV is inverted. On days when the battery does not require a full recharge, the voltage limits are reduced to keep the battery operating in the lower corrosion region. The proposed voltage limits are shown in Table \ref{tab:v_table} compared to the values used by BBOXX and the datasheet recommended values.

An event based approach is implemented for ageing aware voltage control \cite{Collath2022AgingReview}. The maximum number of days between full recharges is 6 to minimise the risk of extensive sulphation and the minimum is 1. A linear scale depending on the proportion of degradation due to corrosion over the previous day is used to determine the number of days between full recharges ($D$) within these limits:
\begin{equation}
\label{vceq}
    D(t) = 1 + 5 \times \frac{C_\text{corr}(t)-C_\text{corr}(t - \Delta t)}{C(t)-C(t - \Delta t) }
\end{equation}

The voltage limits for each time step are dictated by whether a full recharge is required, i.e. if $D$ is greater than the days since the last full recharge, the `full' limits are used, and `partial' limits otherwise.

Using dynamic voltage control means the voltage limits are determined by the batteries' degradation history and can adapt to reduce the dominant ageing mechanism, which changes throughout the lifetime of the battery, rather than assuming static voltage limits will suffice for the whole lifetime \cite{Allahham2022AdaptiveDegradation}.

\begin{table}[]
\caption{Voltage limits as per datasheet reccomendations, the BBOXX systems limits and the proposed limits.}
\centering
\begin{tabularx}{\linewidth}{lXXXX}
\hline
\multirow{2}{*}{}               & \multirow{2}{*}{Datasheet} & \multirow{2}{*}{BBOXX} & \multicolumn{2}{l}{Proposed control} \\
                                &                            &                        & Full     & Partial     \\ \hline
$V_{\text{limit}}$ (V)         & 14.4-15                    & 14.5                   & 14.5        & 13             \\
$V_{\text{float}}$ (V)               & 13.6-13.8                  & 13.5                   & 13.5      & 12.8             \\
$T_{\text{var}}$ (mV/\degree C) & -30                        & 0                      & -30       & -30             \\
\hline
\end{tabularx}
\label{tab:v_table}
\end{table}

\begin{figure}
    \centering
    \includegraphics[width=0.99\linewidth]{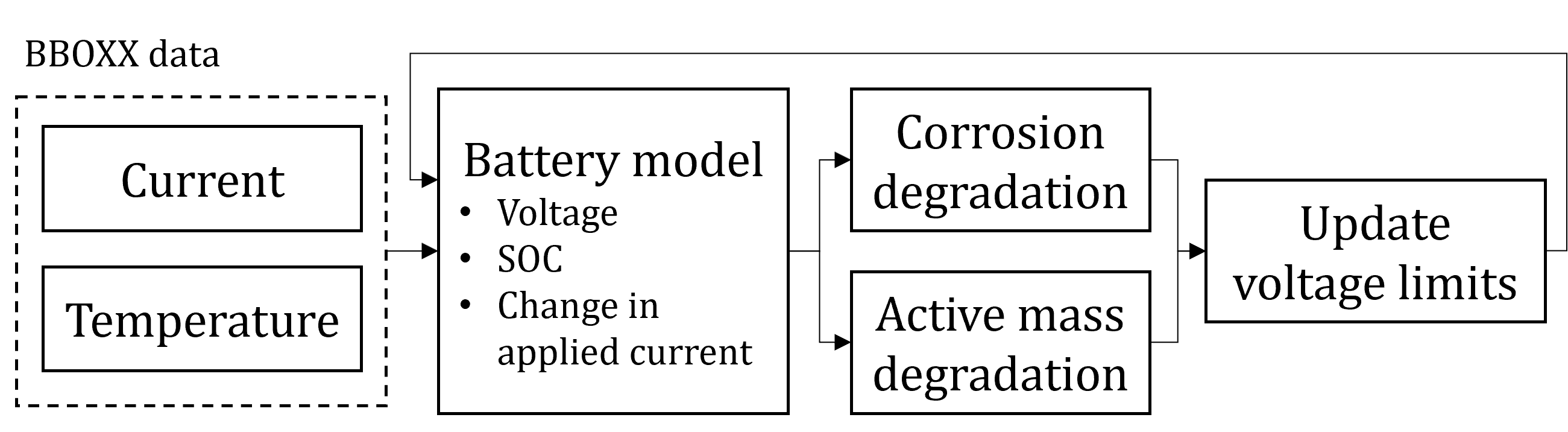}
    \caption{Block diagram for the simulation, combining the BBOXX data, Schiffer battery model and proposed voltage control method.}
    \label{fig:mod_diag}
\end{figure}

The need to vary voltage limits with temperature is clearly demonstrated in literature \cite{Yang2006InfluenceSystems, Bogno2017ImprovementSystems}. A linear approach is considered for this work; each 1\degree C increase in temperature decreases the voltage limits by 30 mV. This is particularly important because the batteries are operating across Africa, where temperatures are often above the recommended operating range of VRLA batteries of 15-25 \degree C (288-298K) \cite{Yang2006InfluenceSystems}. SHS battery charging follows a diurnal pattern, thus the battery is at higher voltages during the hotter periods of the day, highlighting the need for temperature dependant voltage limits.

Fig. \ref{fig:mod_diag} demonstrates how the simulation is run, combining the BBOXX data, battery model (Eqs \ref{batb}-\ref{bate} \& \ref{tot}), corrosion (Eqs \ref{cb}-\ref{ce}), active mass degradation (Eqs \ref{amb}-\ref{ame}) and voltage control method (Eq \ref{vceq}).

\section{Results}
\subsection{Different use cases}
\begin{figure}
    \centering
    \includegraphics[width=0.95\linewidth]{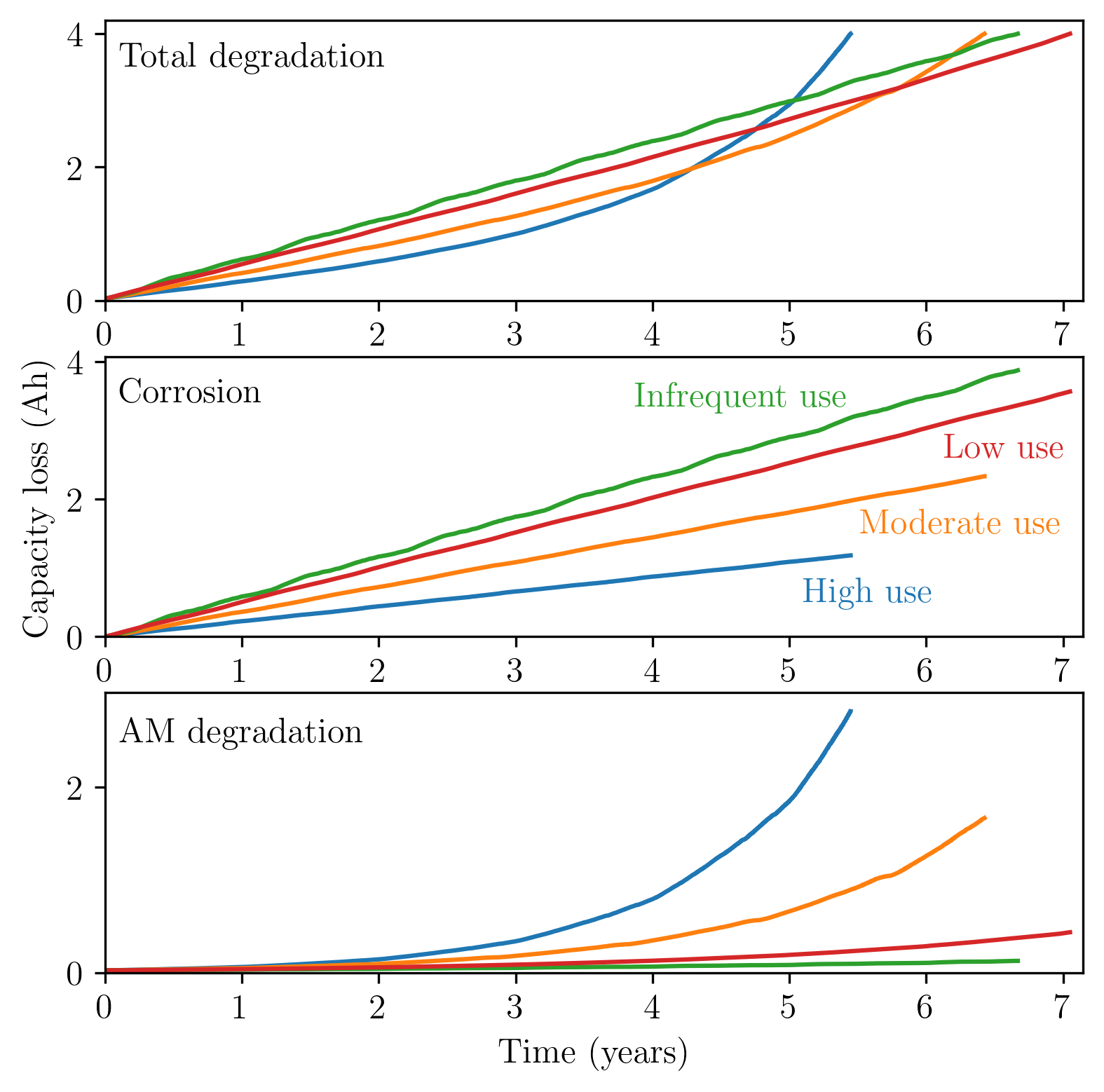}
    \caption{Simulation results until EOL for high use (blue), moderate use (orange), infrequent use (green) and low use (red) showing total degradation and the breakdown of this into corrosion and active mass degradation.}
    \label{fig:cluster_deg}
\end{figure}

The results from running the battery model for the different use cases until EOL (once battery capacity reaches 80\% of nominal capacity) are in Table \ref{tab:res_table} with plots of the degradation mechanisms over the battery lifetimes in Fig. \ref{fig:cluster_deg}.

\begin{table}[]
\caption{Simulation results comparing the lifetime of the battery, full equivalent cycles (FEC) and the percentage of the capacity loss at EOL that is due to corrosion.}
\centering
\begin{tabular}{llll}
\hline
          & Lifetime (years) & FEC before EOL & Corrosion (\%) \\
\hline
High use & 5.4      & 550              & 30                   \\
Moderate use & 6.4      & 489              & 58                   \\
Low use & 7.1      & 327              & 89  \\
Infrequent use & 6.7      & 186              & 97                   \\
\hline
\end{tabular}
\label{tab:res_table}
\end{table}

While the high use case has the shortest lifetime, the greatest number of cycles are completed in this lifetime. This is reflected in reduced corrosion levels compared to other cases, as the increased cycling causes the battery to spend more time at lower voltages.

However, in all other cases corrosion is the dominant ageing mechanism, particularly in low and infrequent use cases. With infrequent use, the prolonged non-use periods mean the battery is kept at float charge for long periods of time. Similarly, low use results in small depths-of-discharge and quick recharges, also resulting in more time at float charge. Keeping VRLA batteries at float charge for prolonged periods is known to cause corrosion \cite{Wong2008ChargeCompensation,Ruetschi2004AgingBatteries}. The detriment of extensively keeping the batteries at float charge is demonstrated by the shorter lifetime of infrequent use compared to low use, even though just over half the number of FEC are completed.

The lifetimes of the batteries in the field do not reach the lifetimes predicted by the simulation. This is likely because the simulation limits are determined by the datasheet details on float and cycle life, which are often not found by systems in the field. Nevertheless, the results are still useful for relative comparisons and insight into the failure modes of the batteries under different use conditions.

\subsection{Proposed voltage control}

\begin{figure}
    \centering
    \includegraphics[width=0.95\linewidth]{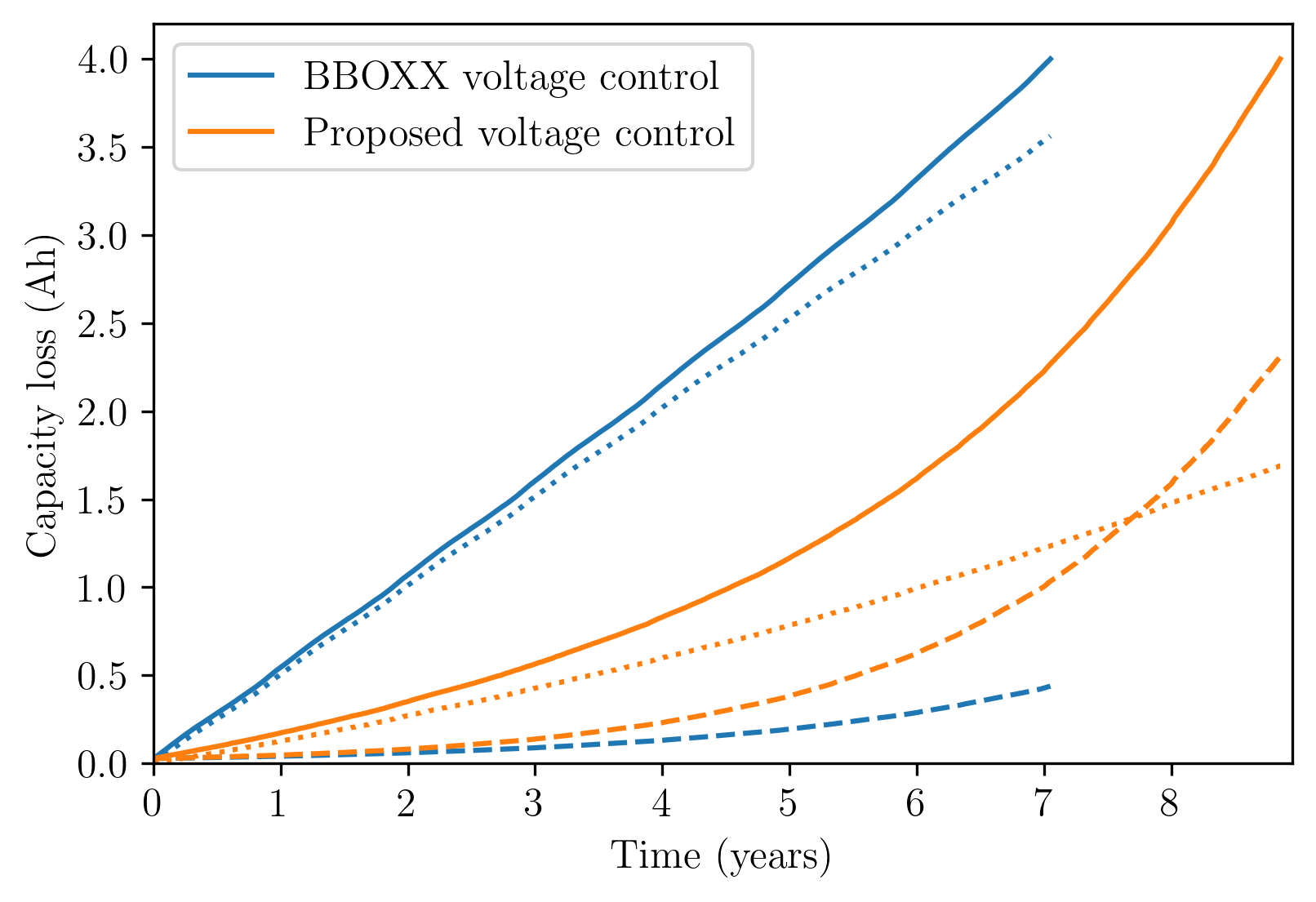}
    \caption{Simulation results until EOL for the BBOXX and proposed voltage control methods. The dotted and dashed lines represent capacity loss due to corrosion and active mass degradation respectively.}
    \label{fig:diff_v}
\end{figure}

\begin{figure}
    \centering
    \includegraphics[width=0.95\linewidth]{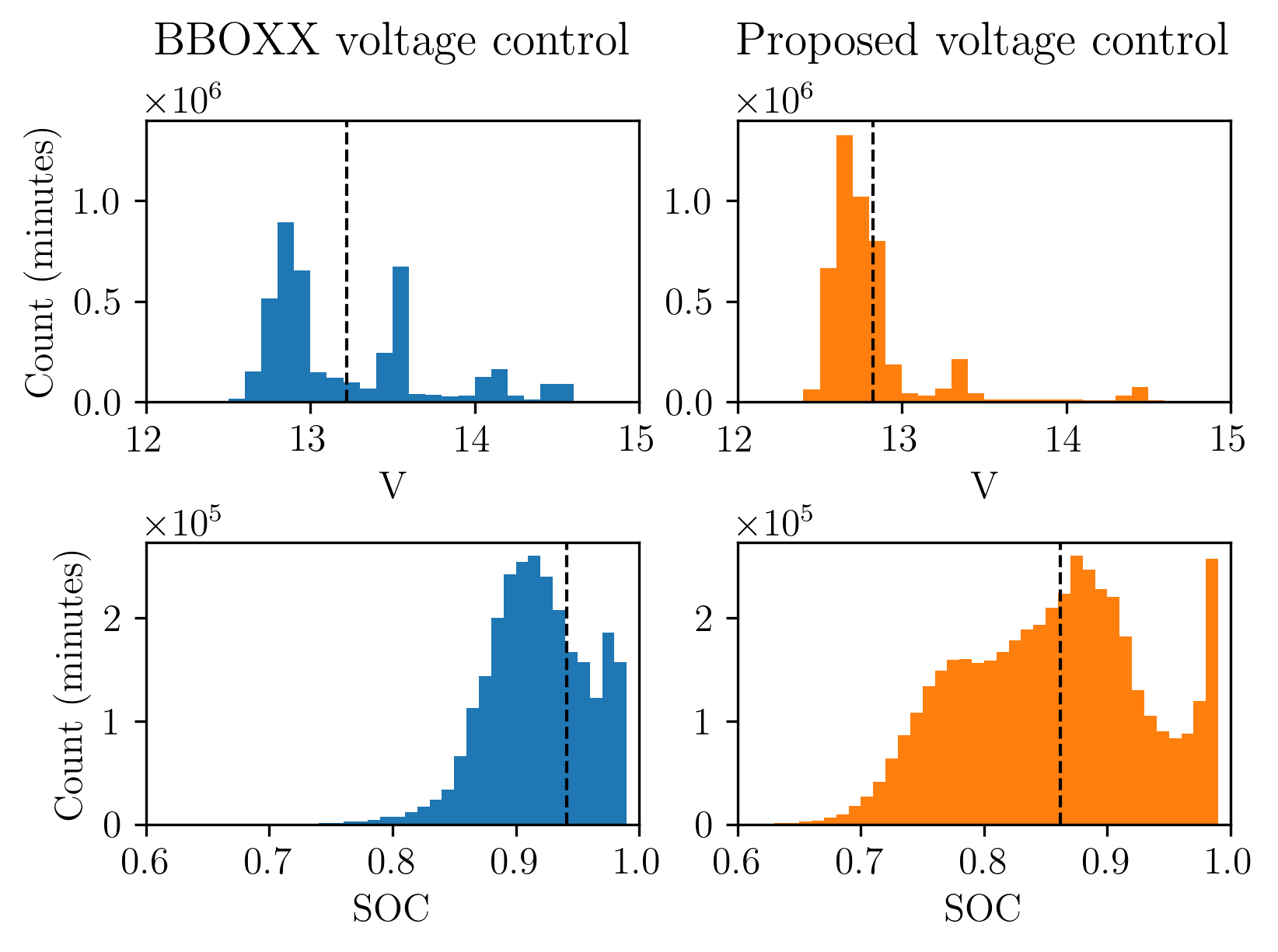}
    \caption{Histograms for time spent in each voltage and SOC region for the low use system under the BBOXX voltage control (blue) and proposed voltage control (orange). The black dashed lines show the mean.}
    \label{fig:soc_hist}
\end{figure}

The dominance of corrosion in the low use case, alongside under-utilisation of the battery and small depth-of-discharge, makes it a suitable choice to try the proposed voltage control methodon (detailed in section \ref{vc}). Fig. \ref{fig:diff_v} shows the difference in degradation for the low use case, with the proposed voltage control methods. It is important to note that the proposed method had no impact on the battery use from the user perspective, and there was no load loss as the SOC never goes below the datasheet reccomended cut-off limit of 50\%. In fact, the SOC remained above 60\% at all times. Fig. \ref{fig:soc_hist} shows the change in time spent in different SOC regions under the proposed voltage control strategy.

The new voltage control method results in a 25\% increase in lifetime. This is predominantly from the significant 47\% decrease in corrosion. However, this comes at the cost of 5 times increase in active mass degradation, due to the increased time between full charges and reaching lower SOC during the cycling. Yet the reduced corrosion, which is modelled as a more linear degradation mechanism than the exponential nature of active mass degradation, means that the battery has a significantly larger capacity under the proposed voltage control strategy at all points in time. At the point of failure under the old voltage control strategy, the new strategy has experienced only 2.26 Ah capacity loss (89\% SOH). Therefore, the proposed strategy doesn't just increase the time before EOL is reached, it also maintains a healthier battery for longer. In addition, active mass degradation caused by increased time between full charges is often due to sulphation. There is evidence in literature that intense overcharges can reduce this sulphation \cite{Wagner2001ChargeApplications}, however this process is not included in the model used in this work.

\section{Conclusion}

This work uses a lead acid battery model from literature combined with field data from SHS across Africa to model how different use cases affect the degradation. Corrosion is found to be the dominant ageing mechanism in all but the highest use case. The different use significantly impacts battery lifetime and FEC completed in this time. The infrequent use case completes fewer cycles whilst also having a shorter lifetime than the low use case, emphasising the problems with simple cycle counting degradation models and the potential benefits from optimised cycling.

To further this work, a dynamic voltage control scheme was proposed, to vary the frequency of full recharging depending on the degradation mechanisms experienced by the battery. The low use case was simulated with this new control scheme, yielding a 25\% improvement in lifetime and FEC while ensuring no loss of load to the user.

\section{Acknowledgements}
The authors thank BBOXX\ for project funding and access to SHS data and EPSRC for project funding.

\small
\bibliography{references}
\bibliographystyle{IEEEtran}

\end{document}